\documentclass[aps,prd,superscriptaddress,letterpaper,preprintnumbers,asmmath,amssymb]{revtex4}
\usepackage{amssymb,amsmath,amsbsy}
\usepackage{mathrsfs}
\usepackage[dvips]{graphics}
\usepackage{epsfig}

\def\be{\begin{eqnarray}}
\def\ee{\end{eqnarray}}
\def\bea{\begin{eqnarray*}}
\def\eea{\end{eqnarray*}}

\def\tilde{\widetilde}

\def\centeron#1#2{{\setbox0=\hbox{#1}\setbox1=\hbox{#2}\ifdim
\wd1>\wd0\kern.5\wd1\kern-.5\wd0\fi
\copy0\kern-.5\wd0\kern-.5\wd1\copy1\ifdim\wd0>\wd1
\kern.5\wd0\kern-.5\wd1\fi}}
\def\ltap{\;\centeron{\raise.35ex\hbox{$<$}}{\lower.65ex\hbox{$\sim$}}\;}
\def\gtap{\;\centeron{\raise.35ex\hbox{$>$}}{\lower.65ex\hbox{$\sim$}}\;}

\newcommand{\newc}{\newcommand}
\newc{\qbar}{{\overline q}}
\newc{\Kahler}{K\"ahler }
\newc{\deltaGS}{\delta_{\rm GS}}

\newcommand{\SmMET}{E_T\hspace{-0.230in}\not\hspace{0.18in}}

\begin{document}
\preprint{
\vbox{\vspace*{2cm}
      \hbox{UCI-TR-2010-24}
      \hbox{October, 2010}
}}
\vspace*{3cm}

\title{Fourth Generation Lepton Sectors with Stable Majorana Neutrinos: From LEP to LHC}
\author{Linda M. Carpenter}

\affiliation{Department of Physics and Astronomy  \\
   University of California Irvine, Irvine, CA U.S.A. \\
   lcarpent@uci.edu  \\
\vspace{1cm}}

\begin{abstract}

I analyze a fourth generation lepton sector in which the lightest particle is a stable Majorana neutrino.  In this scenario fourth generation neutrinos have both a Dirac and Majorana mass, resulting in two Majorana neutrino mass eigenstates.  A reanalysis of LEP's lower mass bound is performed on stable Majorana neutrinos from the Z width and the lower mass bound is loosened.  I also extrapolate LEP's SUSY squark search with a 2 jet plus missing missing energy final state to the production and decay of a pair of heavy Majorana neutrinos; here it is expected that significant regions of the neutrino mass plane may be ruled out.  Finally, a search strategy is proposed for heavy fourth generation neutrino pairs at LHC in the four lepton plus missing energy channel.  Exclusions are set in the neutrino mass plane for 30 $fb^{-1}$ of LHC data at 13 TeV.

\end{abstract}

\maketitle

\section{Introduction}

One of the simplest possibilities for beyond the standard model physics is the existence of a fourth generation of particles.  If a fourth generation exists, its particle masses would naturally lie at the electroweak scale and the leptons would likely be the lightest sector, hence the first to be accessible to current colliders.  Even such a simple extension of the standard model presents rich possibilities for phenomenology, with fourth generation leptons producing signals with many jets, large missing energies, or multi-leptons.  This work addresses a fourth generation lepton sector in which the lightest particle is a stable Majorana neutrino.  I present updated constraints from linear colliders on this scenario, as well as propose possible search strategies at LHC.  Signatures from a fourth generation lepton sector are complex, but also rather challenging for hadron colliders.   Hadron collider seraches will require looking in unusual final states and I present an LHC sensitivity search in a multi-lepton channel.

Recent work has shown that a fourth generation may be consistent with electroweak precision constraints for certain splittings of fourth generation quark masses \cite{He:2001tp,Kribs:2007nz}.  More recent updates have constrained previous work, but still allow a fourth generation at the 95 percent confidence level \cite{Erler:2010sk}\cite{Eberhardt:2010bm}.  Fourth generation quark searches are well underway at current hadron colliders; Tevatron places bounds on fourth generation squarks of of 335 GeV for top type quarks \cite{cdftp} and 385 GeV for bottom type quarks ~\cite{cdfbp}, in addition there have been search strategies proposed at LHC \cite{Ozcan:2008zz,Cakir:2008su,Holdom:2007ap}.

The most general lepton sector of a fourth generation consists of a charged lepton with a Dirac mass, and a neutrino with both a Dirac and Majorana mass.  This means that there are two split Majorana neutrino mass eigenstates.  It is a common assumption that one of the Majorana neutrinos is the lightest fourth generation lepton.  The fourth generation leptons couple to other fourth generation leptons through normal gauge couplings, and they may couple to the Standard Model leptons through gauge couplings suppressed by CKM matrix elements. These matrix elements are generally small, therefore when the heavier fourth generation particles decay they are very likely to do so into a lighter fourth generation state plus a gauge boson. Figure 1 shows possible decay scenarios for fourth generation leptons.  The lightest state may itself decay to SM particles through a nonzero CKM matrix element, or if the neutrino CKM's are zero or very small, the lightest particle may be stable on collider lifetimes.

\begin{figure}[h]
\centerline{\includegraphics[width=6 cm]{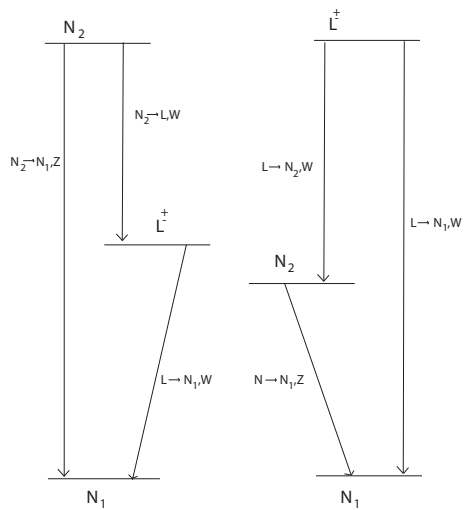}}
\caption{{Decays of Fourth Generation Leptons into Lightest Neutrino. Here L is the charged lepton, and $N_i$ are the neutrinos. }}
\label{fig:2pgm}
\end{figure}

Fourth generation lepton sectors with unstable neutrinos as the lightest particles have been recently studied.  In these cases the lightest Majorana neutrino may decay though a CKM suppressed coupling to a W and standard model lepton, $N_1\rightarrow W\ell$.  Mass bounds in this scenario are all placed by LEP \cite{Achard:2001qw,Abulencia:2007ut}.  The mass of the charged lepton is excluded up to  $\sim$100 GeV.  Mass bounds on pure Majorana or Dirac type neutrinos are 80.5 and 90.3 GeV respectively.  Recent analysis has shown that when neutrinos have both a Dirac and Majorana mass,  the coupling of neutrinos to the Z boson may be reduced by powers of the neutrino mixing angle, hence neutrino pair production is suppressed. In this case, mass bounds on the lightest neutrino may be relaxed as far as 63 GeV \cite{Carpenter:2010dt}.

Unstable neutrinos lead to interesting and easy to find signatures at hadron colliders.  For example, the decay of pairs of the lightest Majorana neutrino leads to signatures with like sign di-leptons in half of all decays; charged leptons cascade decay through the lightest neutrino, and thus also have hard leptons in their decay chain.  Because of this signature, fourth generation leptons- if they exist- may be quite easy to see  \cite{Rajaraman:2010ua}\cite{Rajaraman:2010wk}.  For example by considering $LN_1$ production and looking in the like-sign dilepton channel, charged leptons may be ruled out to 250 GeV by LHC with just 1 $fb^{-1}$ of data at 7 TeV \cite{Carpenter:2010bs}.

The case where the lightest Majorana neutrino is stable on collider lifetimes is much less studied and much less constrained.  While the bound on a charged lepton in this scenario is still set by LEP around 100 GeV, the bound on stable neutrinos themselves is extremely low.  In fact the only bound on the neutrino sector comes from LEP's measurement of the invisible Z width.  Contributions to the invisible Z width from any new physics including stable neutrinos must be under 21 MeV.  This translated to a Majorana neutrino mass bound of just 39.5 GeV \cite{pdg}\cite{Abreu:1991pr}.

In the  more general case  where there are two Majorana neutrinos, the Z pole limits must be revised.  Again the coupling of neutrinos to the Z boson is reduced by powers of the neutrino mixing angle, therefore there are regions of large mixing in the $N_1 N_2$ mass plane where the invisible Z width is reduced.  I recalculate the lower neutrino mass bound from the Z width and show it may be significantly lowered to 33.5 GeV.

Though the Z pole neutrino mass constraint may be lowered, existing SUSY searches at LEP have the possibility to severely constrain fourth generation parameter space in the case that there are two Majorana neutrinos.   There are many events at LEP in which a heavy and light neutrino pair are produced through the process $e^{+} e^{-} \rightarrow N_2 N_1 \rightarrow Z N_1 N_1 \rightarrow 2j +\SmMET $.  The final state is two jets plus missing energy, which was a channel searched for by LEP's SUSY squark search \cite{Achard:2003ge}.  In this work I attempt to constrain neutrino mass parameter space by applying the results of LEP's degenerate squark search to fourth generation neutrinos and I find large exclusions over the $N_1, N_2$ mass plane.

Unlike the situation at LEP, searches for fourth generation leptons in the case that a stable Majorana neutrino is the lightest particle are quite challenging for hadron colliders.  Decays of pairs of particles proceed to stable neutrinos plus gauge bosons; for example $L N_2 \rightarrow WZ N_1 N_1$ or  $N_2 N_2 \rightarrow ZZ N_1 N_1$. Due to the gauge boson branching ratios, most final states consist of jets plus missing energy. However, these signals will be quite hard to for LHC to resolve as - unlike gluinos which have a similar signals - these leptons only have an electroweak production cross section.  In order to search for this scenario, one needs to choose  more distinctive signatures.

Fourth generation leptons, however, are capable of producing more peculiar signals than just jets.  I propose as an example a search for pairs of the heavier neutrino decaying to the lightest neutrino state, $N_2 N_2 \rightarrow ZZ N_1 N_1$ in the four leptons plus missing energy channel.  In this case I consider the leptonic decay of both Z's.  Though the branching fraction is small, the background is nearly zero.  I present exclusions in the $N_1, N_2$ mass plane for a 13 TeV run of LHC with $30 fb^{-1}$ of data.

This work proceeds as follows, Section 2 reviews the mass formalism for fourth generation neutrinos.  Section 3 addresses the LEP Z pole constraint and probable exclusions in the $N_2 N_1$ mass plane from an existing jets plus missing energy search at LEP.  Section 4 present a sensitivity study for heavy neutrino pairs at LHC in the $4\ell + \SmMET $ channel, and Section 5 concludes.



\section{Review of Fourth Generation Lepton Masses and Couplings}

  Consider a fourth generation lepton system with the most general possible masses.  In this sector fourth generation charged leptons have a Dirac mass, while fourth generation neutrinos have both a Dirac and a Majorana mass.

The neutrino mass terms maybe written
\begin{eqnarray}
{\cal L}_m=-{1\over 2}\overline{(Q_R^c
N_R^c)}\left(\begin{array}{cc}0&m_D\\m_d&M\end{array}\right)
\left(\begin{array}{c}Q_R\\ N_R\end{array}\right)+h.c.
\end{eqnarray}
Diagonalizing the mass matrix, one finds two neutrino mass eigenstates $N_1$ and $N_2.$  In the mass eigenstate basis the Lagrangian becomes

\be
L_m={m_1} \bar{N}_{1}N_{1}+ m_2 \bar{N}_{2}N_{2}+ \sum_{i=1}^4m_{e}^{i}\bar{L}_i E_{Ri}
\ee
where $N_i$ are the neutrinos and L and E are the charged leptons.  The neutrino masses are given by
 \be
 \nonumber m_1=-(M/2)+ \sqrt{m_D^2+{M^2/4}}; \\ m_2=-(M/2)- \sqrt{m_D^2+{M^2/4}}
\ee
while the mass eigenstates are given by the neutrino mixing angle;
\be
N_{1}=\cos\theta\nu_{4L}+\sin\theta\nu_{R}^c ;   N_{2}=\cos\theta\nu_{R}^c-\sin\theta\nu_{4L}
\ee
with
\be
\nonumber \tan\theta=m_1/m_D
\ee
Notice that in the Dirac limit, $m_1$ is equal to $m_2$, in the majorana limit, $m_2$ is much bigger than $m_1$.

The coupling of the leptons to gauge bosons is given by

\be
{\cal L}=gZ_\mu J^{\mu}+(gW_\mu^+ J^{\mu +} + c.c)
\ee
  with
\be
 J^{\mu }
= {1\over 2\cos\theta_W}(-c^2_\theta\bar N_1\gamma^\mu \gamma^5N_1
-2is_\theta c_\theta\bar N_1\gamma^\mu N_2-s^2_\theta\bar N_2 \gamma^\mu\gamma^5N_2))
\\
J^{\mu +}=
c_i\overline{(c_\theta N_1-i s_\theta N_2)}\gamma^\mu l^i_L~ +{1\over\sqrt{2}} \overline{(c_\theta N_1-i s_\theta N_2)}\gamma^\mu L~~~~~~~~~~~~~~~~~~~
\ee

Notice that the coupling of the leptons to the gauge boson varies with mixing angle.

\section{Neutrino Mass Constraints from LEP}

I now elaborate the relevant neutrino mass constraints from LEP.  Here, pair production of neutrinos occurs through a Z boson thus the relevant processes for studying the neutrino sector are $e^{+}e^{-} \rightarrow Z\rightarrow N_i N_j$.  The lightest neutrino state is of course stable on collider lifetimes, the heavy state, however, promptly decays to the lightest neutrino through an off or on shell Z boson.  One must determine the allowed masses of the neutrinos in this scenario.

\subsection{Z Pole Constraints}

The current mass bound on a stable Majorana neutrino comes only from the invisible width of the Z boson.  To two sigma, the Z width may be corrected by 21 MeV in the invisible channel and 33MeV overall \cite{pdg}\cite{Abreu:1991pr}.  If a stable neutrino exists below half of the Z mass, then the Z may decay invisibly to neutrino pairs, increasing its invisible width.  In the case that a single stable Majorana neutrino exists, this translates to a neutrino mass exclusion of 39.5 GeV.  In the case that there are two Majorana neutrinos, the relevant contributions to the Z width come from three processes
\be
Z\rightarrow N_1 N_1 \rightarrow \SmMET \\
Z\rightarrow N_1 N_2\rightarrow \SmMET + 2j,2\ell \\
Z\rightarrow N_2 N_2\rightarrow \SmMET + 4j,4\ell, 2j+ 2\ell
\ee

The first process contributes to the invisible width of the Z, while the last two contribute only to the total width.  If the Z coupling to the lightest stable neutrino is decreased, then the mass bound from the invisible Z width may be loosened; in the case that there is a mixing between neutrino states this is precisely what occurs.   One sees from equation 6 that the the stable neutrino coupling to Z decreases as the neutrino mixing angle increases.  The suppression of the Z coupling goes like square power of cos$\theta$, hence the stable neutrino mass bound may be lowered in the region of parameter space where the mixing is large.  In order to minimize the invisible width, one wants to make the cosine of the neutrino mixing angle as small as possible, for fixed $m_1$ this means lowering $m_2$ thus compressing the neutrino masses.

If the $N_2$ state is also kinematically accessible, then the Z may decay to it as well.  If Z decays to $N_2$ pairs or to $N_1 N_2$ there will be a contribution to the total, but not invisible, Z width. The lighter $N_2$ is, the larger the contribution to the total Z width.  There is then some tension between bounds from the invisible and total width of the Z as one hits the constraint from the total width as the neutrino masses are compressed.  The excluded regions in the $N_1$ $N_2$ mass plane due to Z width measurements are shown in Figure 2.  If one considers only constraints from the Z width the minimum allowed $N_1$ mass is relaxed to 33.5 GeV.  Extra limits may be placed on this situation however.  It may be possible that astrophysical measurements from neutrino anihilation constraint lighter neutrino masses in this scenario, though no calculation has been made for split mass Majorana neutrinos, see for example \cite{Fargion:1994me}.  One will see in the next section that existing LEP SUSY searches greatly constrain the parameter space.




\subsection{LEP SUSY Search Constraints }

As one sees from Eqn. 8, the production of neutrino pairs where a heavy neutrino is present resulting from the process $e^{+}e^{-}\rightarrow Z\rightarrow N_i N_j$  most likely results in decays containing jets plus missing energy. This is a common final state in SUSY searches and one that was looked for by LEP's squark search. In the SUSY search the process under consideration was $e^{+}e^{-}\rightarrow \tilde{q} \tilde{\overline{q}} \rightarrow  qq+\chi_0\chi_0$,  where the squarks decayed to a quark and stable neutralino.  This search excluded degenerate squark masses up to 99.9 GeV \cite{Achard:2003ge}, which is very near LEP's kinematic limit.  One may then reasonably expect large exclusions in searches with a similar final state, and one sees that the production of $N_1 N_2$ and $N_2 N_2$ results in similar or identical final states. What is more, the production cross section of sfermions at LEP should be about the same as for heavy neutrino pairs, so one would expect similar sensitivities.  Though no specific search has been done for neutrino pairs, probable limits on the exclusion in the $N_1 N_2$ mass plane may be placed from this existing SUSY search.

In this analysis $N_1 N_2$ and $N_2 N_2$ pairs were generated using MADGRAPH \cite{Alwall:2007st}. The events were decayed with BRIDGE \cite{Meade:2007js} and showered with PYTHIA \cite{pythia}.  LEP's acoplanar jet preselection cuts were implemented \cite{Achard:2003ge}\cite{Acciarri:1999kk}.   The acoplanar jet search found 3110 events with $3514 \pm 18$ expected from the di-photon background.    The pre-selection looked for 2 jet events with the following cuts,

\begin{itemize}
\item{at least 4 tracks }
\item{$5<E_{vis}<$150 GeV }
\item{$E_T >$10 GeV  in forward calorimeters }
\item{energy in 30 degree cone around beam pipe $<.25 E_{vis}$ }
\item{missing $p_T >$2 GeV }
\item{$sin\theta_{p_{miss}}>$.2 }
\end{itemize}

\begin{figure}[h]
\centerline{\includegraphics[width=6 cm]{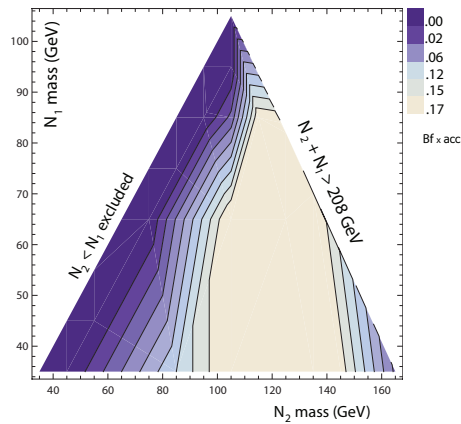}}
\caption{{Plot of search acceptance for acoplar jets in the $N_1$ vs $N_2$ mass plane. }}
\label{fig:2pgm}
\end{figure}

\begin{figure}[h]
\centerline{\includegraphics[width=10 cm]{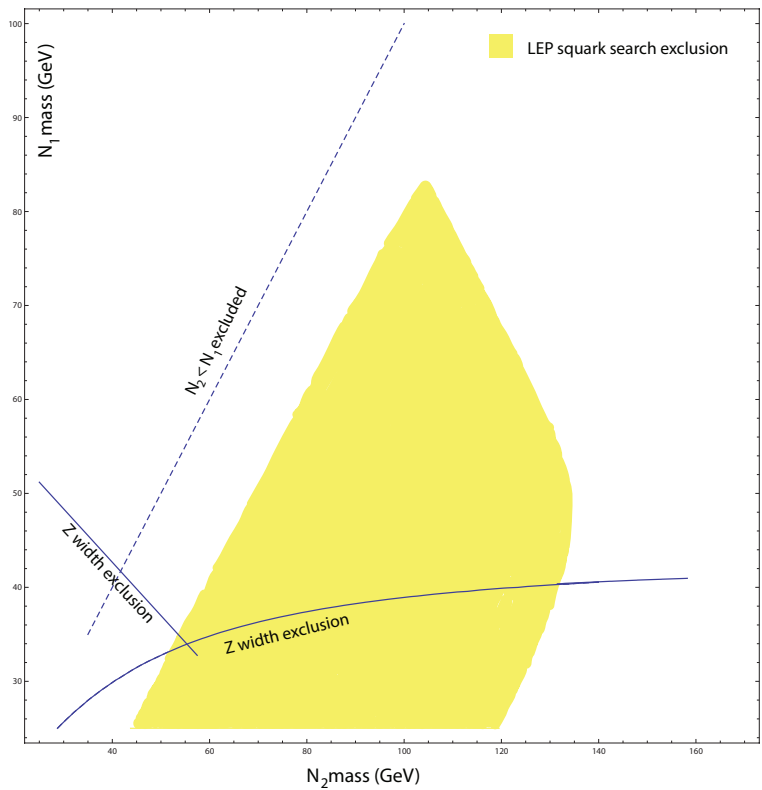}}
\caption{{Plot of allowed $N_1$ vs $N_2$ mass.  The yellow region shows LEP's likely exclusion from SUSY searches.  The solid lines are Z pole exclusions.}}
\label{fig:2pgm}
\end{figure}

These cuts were implemented as the basis of this exclusion.  Over most of the mass plane, calculated efficiencies range up to .19, except where the masses of $N_2$ and $N_1$ are close together, is which case the jets become very soft and efficiency is lost.  Figure 3 shows efficiencies over the $N_1$ $N_2$ mass plane.   A slight loss of efficiency occurs for regions where $N_2$ is much more massive than $N_1$.  In this case the decay products of $N_2$ are boosted and the two jets from $N_2$ decay become less distinguishable.  The expected exclusion in the $N_1 N_2$ mass plane from a jets plus missing energy search is shown in Fig 3.  The exclusion from the Z width is given on the same plot.  All exclusions are given at the 95 percent confidence level.  One sees that unless the neutrino masses are degenerate, the heavy neutrinos are excluded for most of the mass plane up to LEP's kinematic limit.  $N_2$ exclusions in the non-degenerate mass region range up to at least 100 GeV, and in some cases 130 GeV.   In this same region $N_1$ masses may be excluded at least up to 80 GeV.  In the region of degenerate neutrino masses, one finds the minimum allowed $N_1$ mass set by the Z width constraints at 35 GeV.
\begin{figure}[t]
\begin{center}
\includegraphics[width=0.32\linewidth]{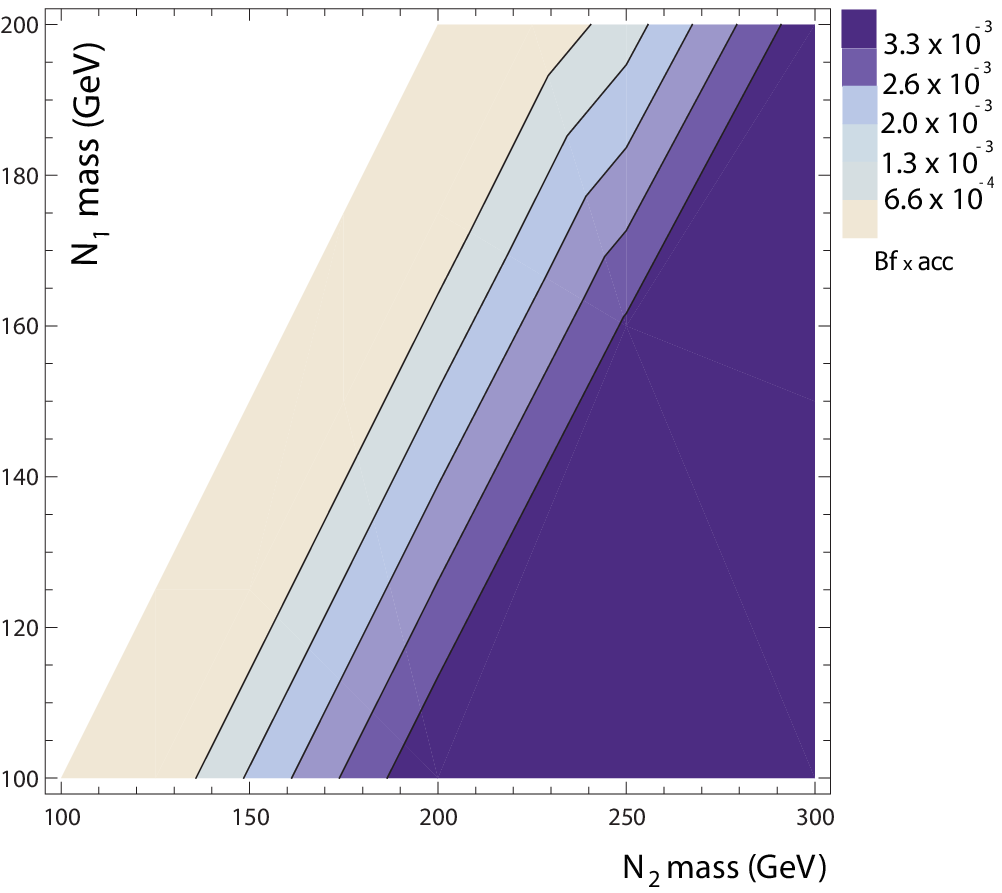}
\includegraphics[width=0.3\linewidth]{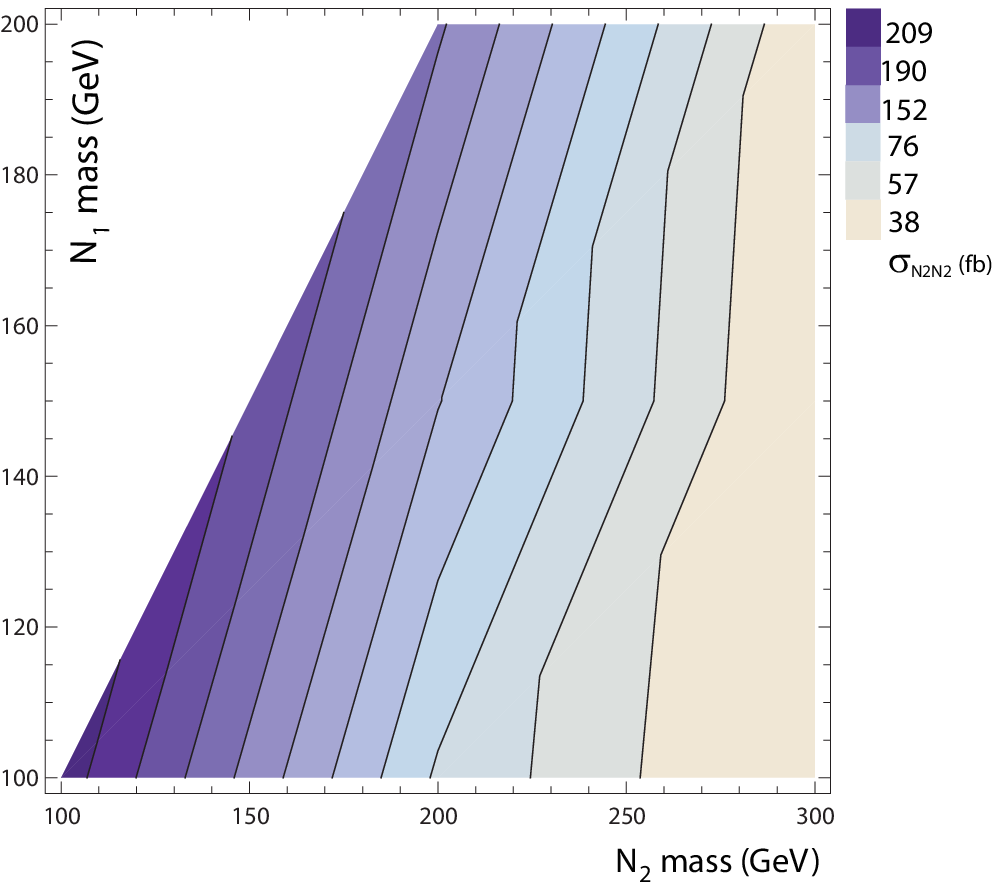}
\caption{  Left, search acceptance over the $N_2$, $N_1$ mass plane for a 4$\ell$ + $\SmMET$ search. Right, Production cross section of $N_2$ pairs over the mass plane.}
\label{fig:reco}
\end{center}
\end{figure}

\subsection{Hadron Collider Searches}

Moving on from current constraints, one must find channels in which fourth generation leptons may be discovered or ruled out at current hadron colliders.  The most obvious processes to look for at hadron colliders would be the most probable ones, which means jets plus missing energy final states.  One may ask if current search strategies have a hope of finding heavy leptons in these channels.  Both Tevatron and LHC searches for squarks and gluinos would be relevant to the $4j+\SmMET$ and $2j+\SmMET$ decay channels of neutrino or charged lepton pairs.   However these are in fact not good channels for looking for fourth generation lepton sectors with stable lightest neutrinos.  I will illustrate this by considering Tevatron's current SUSY searches.

Tevatron's gluino search looks for the pair production and decay of gluinos decaying to jets and a stable neutralino.  The process under consideration is $p\overline{p}\rightarrow \tilde{g} \tilde{g} \rightarrow  qq\tilde{q}^{*}\tilde{q}^{*}\rightarrow qqqq\chi_0\chi_0$, which is a 4 jet plus missing energy final state.  The current gluino exclusion is given by PDG at 308 GeV \cite{pdg} and one might think that pairs of fourth generation leptons which decay to an identical final state might also be excluded by this search.  This search and any like it, however, do not have bearing for the fourth generation scenario. Tevatron looks for gluinos which have a strong production cross section, of order 10 pb .  In the case of fourth generation leptons however, the production is through electroweak processes and is much lower, expected to be only of order 100fb.  Anything with electroweak production cross section would require a search with extremely distinguishing cuts to separate it from the large QCD background.
For example current searches for gluinos only make large exclusions for gluino masses in the MSUGRA scenario in which the gluino is much heavier than the neutralino it is decaying into.  In MSUGRA in fact $m_g:m_B$ is $6:1$.  This search relies on a very large missing energy cut and very hard jets.  As the ratio of gluino to neutralino mass is suppressed one must use smaller cuts in analyses and the gluino exclusion loosens. Sensitivities have been estimated for gluino searches which rely on lesser cuts on missing energies; these place less stringent bounds on gluino masses\cite{Alwall:2008ve} and can only claim to exclude gluino masses above 120 Gev so, a much weaker bound.  When studying the case of pair production and decay of weak scale neutrinos, one expects that there will be no large cuts on missing energy or on jet $p_T$, therefore one already faces weaker exclusion even if the production cross section was very large, however the production cross section is also orders of magnitude below that of gluinos. Thus the jets plus missing energy signal from fourth generation leptons will be entirely swallowed by the background.

For this reason it is unlikely that Tevatron would be able to find pair produced neutrinos, or a charged lepton and neutrino pair using this strategy. At LHC, the situation is identical.  The production cross section for a 100 GeV heavy neutrino pair at LHC, about 200 fb, is so small that is equal to that of a pair of 1.5 TeV gluinos.  The heavy gluinos however, would be differentiable from background through the use of extremely hard cuts. Pairs of weak scale neutrinos would not have extremely energetic jets and huge missing energy to distinguish them from background.  Hence, one needs more distinctive signals with low backgrounds in order to find the fourth generation lepton sector in this scenario at hadron colliders.  One must turn to other channels.  A possible set of signals for extracting a fourth generation lepton sector at LHC are those with multiple leptons.  In this case one must exchange branching fraction for clean final states with low background.  Though we will only have the Tevatron a few more years, one can count on LHC's eventual large luminosity to make such states visible.

In this case I have chosen to do a sensitivity analysis for finding fourth generation neutrinos in the 4 leptons plus missing energy final state.  A quad-lepton signal follows from the pair production and decay of heavy neutrinos $p p\rightarrow Z \rightarrow N_2 N_2\rightarrow N_1 N_1 ZZ$.  In this case both Z's decay leptonically, $N_2 N_2\rightarrow N_1 N_1 ZZ\rightarrow \ell\ell\ell\ell N_1 N_1$, thus the final state is $4\ell + \SmMET$.  Though the branching fraction is low, this signal quite a spectacular and in addition, the background is almost non-existent.

One may do a sensitivity analysis for excluding $N_2$ pairs using the $4\ell + \SmMET$ final state.  The following cuts are used in this analysis
\begin{itemize}
\item{4 isolated leptons}
\item{Most energetic lepton $p_T > 15$ GeV, other two lepton $p_T> 5GeV$}
\item{$E_{miss}>50 GeV$}
\item{jet veto; $N_{jet} < 2$}

\end{itemize}

\begin{figure}[h]
\centerline{\includegraphics[width=8 cm]{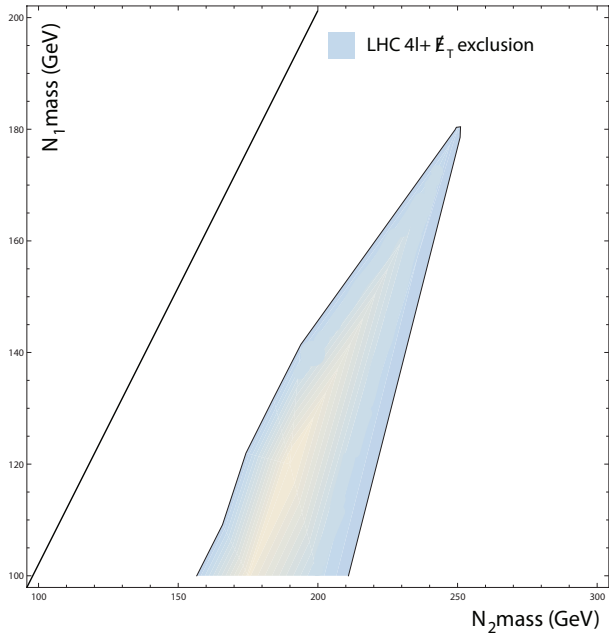}}
\caption{{Plot of allowed $N_1$ vs $N_2$ mass.  The shaded region is exclusion to 95 percent confidence of an LHC 4$\ell+\SmMET$ search. }}
\label{fig:2pgm}
\end{figure}

Events were generated in MADGRAPH \cite{Alwall:2007st}, decayed using BRIDGE \cite{Meade:2007js} and showered through PYTHIA \cite{pythia}.  The acceptances  were calculated using the above cuts. Both the acceptances and the $N_2$ pair production cross sections are shown in Fig 4.  One sees that the search efficiency is highest when $N_2$ is much heavier than $N_1$ and it falls when the $N_2$ mass falls as resultant leptons become soft.  The total production cross section of course falls as the $N_2$ mass is increased.  The background was calculated from ZZ and $\gamma Z$ production and no events were found.  The main factor in background rejection was the size of the missing energy cut. An exclusion plot is given in Fig 5 in the $N_1 N_2$ mass plane with 30 $fb^{-1}$ of data and at p5 percent confidence.  Exclusion of $N_1$ masses range from 100 to 180 GeV, while exclusions of $N_2$ masses range from 150 to 250 GeV.

\section{Conclusions}

Constraints on fourth generation lepton sectors where the lightest Majorana neutrino is stable were analyzed.  Neutrino mass constraints from the Z-width were reanalyzed considering that neutrinos may have a Majorana as well as a Dirac mass.  In addition, mass bounds in the neutrino mass plane were set in light of LEP's acoplanar jet plus missing energy search. Finally a sensitivity analysis was performed for a pair of heavy neutrinos decaying to four leptons plus missing energy at LHC.  Possible exclusions were made in the $N_1$ $N_2$ mass plane with 30 $fb^{-1}$ of data at 13 TeV.

Searching for fourth generation leptons in this scenario is quite challenging for hadron colliders even though the production cross sections for leptons may be quite high.  Finding the charged leptons presents a particular problem.  For example, production of an $LN_1$ pair from a W is quite hight at LHC.  In the case of an unstable $N_1$ state, this process proved quite easy to find at LHC by looking in the like sign di-lepton channel.  However, if the $N_1$ state is stable, then the decay process $L N_1\rightarrow WN_1 N_1$  becomes lost in the W background, which at LHC is very large.  Finding this process in the jets plus missing energy channel is again unlikely due to the overwhelming QCD background.

However there are other distinctive channels to look in for charged leptons.  For example, one may produce $pp\rightarrow W\rightarrow LN_2$.  This may decay to $LN_2\rightarrow WZN_1N_1\rightarrow 2\ell+2j+\SmMET$.  If this signal may be distinguished from the $t \overline{t}$  background , perhaps with a b veto, this may be an interesting channel to look for new physics in.  If one also considers cascade decays of charged leptons through the heavy neutrino state there may be even more unusual signals to analyze.

There may also be a different kinds of searches possible at LHC which place constraints on stable neutrinos.   Barring all else, proposals have been made to search for stable uncharged weakly interacting matter at LHC by measuring recoils \cite{Goodman:2010yf}.  These searches were proposed to look for Dark Matter candidates and claim to be especially sensitive to light weakly interacting particles.  Such a search should be able to place bounds directly on a stable fourth generation neutrino mass.


{\bf Acknowledgments}

This work was supported in part by NSF grant number PHY-0653656. Thanks to Daniel Whiteson and Arvind Rajaraman for helpful discussions.

\end{document}